# Observing the Effects of Legacy 10G on 100G Channels in Dispersion Managed Optical Systems

**Emanuele Virgillito[1], Andrea Castoldi[3], Stefano Straullu[2], Silvio Abrate[2], Rosanna Pastorelli[3] and Vittorio Curri[1]**

[1]DET, Politecnico di Torino, C.so Duca degli Abruzzi 24, 10129 Torino, Italy
[2] LINKS Foundation, Via Pier Carlo Boggio 61, 10138 Torino, Italy
[3]SM-Optics, Via John Fitzgerald Kennedy, 2, 20871, Vimercate, Italy
*emanuele.virgillito@polito.it*

**Abstract:** We show that 10G channels generate both amplitude and phase noise on 100G channels. Amplitude noise can be managed as the ASE and NLI noise, while the DSP robustness to the phase noise sets the guard-band.

## 1. Introduction

In the market of back-bone networks, the bulk of the optical channels is dominated by coherent transceivers deploying 100G and beyond. On the contrary, access and metro segments still rely on 10G transceivers based on the Intensity Modulation Direct Detection (IMDD) and operated on Dispersion Managed (DM) optical line systems (OLSs), because of large CAPEX savings. In particular, the upgrade of legacy but still widespread Synchronous Digital Hierarchy (SDH) networks to simpler WDM architectures is a large market potential for this technology, because a sudden and complete migration to coherent transmission would not be effective from a cost-advantage point of view. In the meantime, these technologies are evolving, targeting cost reduction also thanks to the extensive use of integrated photonics. So, a medium-term realistic scenario will include DM OLSs operated by mixed 10G/100G transceivers.

It is well known that coherent transmission technologies are impaired by DM OLSs [1] and that joint 10G/100G transmission requires a guard-band to avoid large penalties on the 100G comb [2]. So, a realistic transmission scenario is the one depicted in Fig. 1: a DM OLS will deploy mixed 10G/100G transmission on different portions of the exploited bandwidth, separated by a guard-band. So far, as the analysis on the impairments of an IMDD 10G channel comb on 100G channels has not yet been assessed, engineering rules for the guard-band definition are still missing. It is of paramount importance to separate the effects that are generating amplitude disturbance similar to the nonlinear interference (NLI) generated by 100G-to-100G interaction from the ones that are inducing phase noise. The effects of amplitude disturbances on DSP-based coherent receivers is well known and can be managed and put under control by their inclusion in the signal-to-noise ratio (SNR) together with the ASE noise and the NLI. In this case, the use of the SNR as unique quality-of-transmission (QoT) figure holds. More delicate is the effect of phase noise possibly generated by 10G channels: it is not manageable within the SNR and the impairment it generates depends on the implementation of the DSP, mainly at the carrier-phase-estimation (CPE) stage. Moreover, 10G channels are polarized, so, their impairments will be affected by rotation of the PMD principal axes. Thus, all the impairments must be evaluated in the worst-case scenario, corresponding to polarization-aligned 10G channel combs, to avoid random out-of-services due to cycle-slips.

In this work, we observe via simulation the effects of 10G interfering channels on a QPSK probe. We study the effect on one polarization axis to observe the worst-case effects. We show that the impairments arise from cross-phase modulation (XPM) and four-wave mixing (FWM). FWM generates only amplitude noise, while XPM generates both amplitude and phase noise. Also, the phase noise becomes dominant in case of DM OLSs with a small amount of undercompensated chromatic dispersion (CD) at each span, which is a typical scenario in metro WDM systems. Considering an 11-channel 10G comb, we show the overall effect on a QPSK probe focus on characterizing the phase noise as statistics and spectral occupation as it is the main phenomenon to be considered in defining the guard-band.

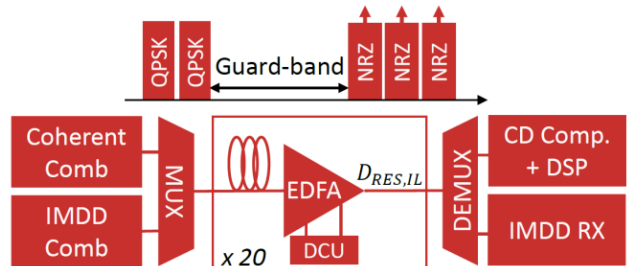

Fig. 1. Scenario for mixed 10G/100G transmission

## 2. Observing the effects of IMDD modulation on PM-QPSK channels

We observe via simulation the effects of IMDD on a QPSK on DM OLS. In such a scenario, non-linear impairments can be classified following the taxonomy for legacy IMDD systems. Self-phase modulation (SPM) is neglected, being a single channel impairment, which involves the coherent probe only; whereas, we focus on XPM and FWM originated by the IMDD pumps on the QPSK probe. Since XPM behaves as a modulation of the probe phase proportional to the pumps power, it can be modeled as a complex multiplicative noise $\rho(t)$ originated by $N_p$ IMDD pumps on the coherent probe, as proposed in [3]. The FWM portion falling on the probe bandwidth is instead an additive Gaussian noise depending only on the IMDD pumps power and fiber physical parameters. Thus, the coherent probe field $a_{RX}$ at the receiver, after CD compensation and before the DSP can be written as Eq. 1.

$$a_{RX}(t) = \rho(t)a_{TX}(t) + n_{FWM}(t) \quad (1)$$

$$\rho(t) = n(t) \cdot e^{j\phi(t)} \quad (2)$$

Here, $a_{TX}$ is the coherent probe field at the transmitter output, $n_{FWM}(t)$ is the FWM disturbance falling within the probe bandwidth. For the XPM noise $\rho(t)$, $n(t)$ is the amplitude noise originated by the phase-to-amplitude conversion due to CD, $\phi(t)$ is responsible for the phase noise on the coherent signal. Also, assuming that superposition principle holds, the cumulative XPM noise $\rho(t)$ is obtained as the product of the $\rho_i(t)$, $i = 1 \dots N_p$ generated by the single IMDD pumps. Hence, $\rho(t) = \prod \rho_i(t)$ and for the single components Eq. 3 holds.

$$n(t) = \prod_{i=1}^{N_p} n_i(t), \quad \phi(t) = \sum_{i=1}^{N_p} \phi_i(t) \quad (3)$$

In order to observe the effects of IMDD on QPSK, we have simulated the propagation of a single QPSK probe propagating together with several 10 Gbps IMDD pumps over a 20-span SMF DM link (Fig. 1), also looking at the impact of different dispersion maps by varying the inline residual dispersion $D_{RES,IL}$ at the end of each span. Actual span lengths have been randomized with respect to the average 50 km to avoid unrealistic FWM resonances. Accumulated CD is fully recovered at the end of the link as it is done by coherent receivers. SPM is avoided by keeping the probe power sufficiently low (-20 dBm). In order to observe XPM effects, we performed an extensive set of pump & probe simulations - i.e., by propagating a coherent probe and one IMDD pump – varying $D_{RES,IL}$ from 0 ps/nm (full CD compensation at the end of each span) to the Uncompensated Transmission (UT), the pump power $P_{pump}$ in the -1 to 4 dBm range and the pump-probe spacing $\Delta f$ in the 50GHz frequency grid until 1 THz spacing. Simulations were done using the FFSS split-step library [4]. Pump & probe configurations ensure that no FWM noise falls on top of the coherent probe; hence, the XPM noise contribution of the $i$-th pump alone $\rho_i(t)$ is obtained from

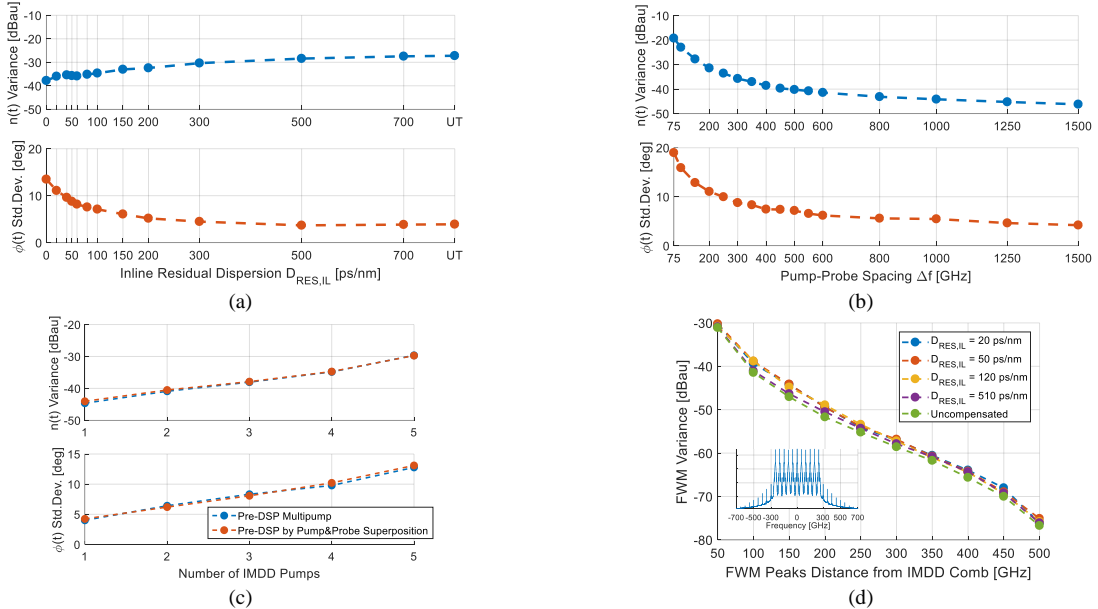

Fig.2. Amplitude noise variance and phase noise standard deviation obtained with 20 span SMF pump & probe simulation varying (a) the dispersion map ($D_{RES,IL}$) and pump-probe frequency spacing $\Delta f$ at $P_{pump} = 1$ dBm. (c) Effects superposition test for XPM effects at $D_{RES,IL} = 50$ ps/nm. (d) Variance of FWM peaks arising from 11 IMDD channels at $P_{pump} = 1$ dBm for various dispersion maps.

Eq. 1 as the ratio between the received and transmitted coherent signal. Then, the amplitude noise $n_i(t)$ variance and the phase noise $\phi_i(t)$ standard deviation have been computed. Fig. 2 shows the most interesting results. As we approach the UT, amplitude noise becomes more significant with respect to phase noise (Fig. 2a) that is in accordance with the existing models for multilevel modulation formats with UT. Also, the entity of the impairment decreases by increasing the frequency separation (Fig. 2b). In Fig. 2c we show that the additivity hypothesis for XPM noise holds by comparing the sum of the pump & probe contributions with proper multipump simulations statistics. Furthermore, we have studied the FWM noise generation of a comb of IMDD channels by turning off the coherent probe. Fig. 2d plots the side FWM peaks variances generated by 11 pumps, with the main finding that the FWM generation is substantially independent of the dispersion map: it is indeed determined only by fiber parameters, the pump powers and their aggregate spectral occupancy. However, these metrics are here not comparable with XPM metrics because of the different nature of the two noises.

## 3. Setting the guard-band

The developed methodology provides a powerful engineering tool to evaluate the entity of the XPM induced phase noise. This acquires importance in the 10G/100G coexistence perspective since the typical dispersion map for DM links using a small under-compensation shows a significant amount of phase noise. From the system planning and orchestration point of view, the amplitude XPM noise can be regarded as another source of additive noise and mitigated by power optimization strategies accounting for it in SNR computation. Phase noise is instead a serious impairment for the DSP processing that can be kept under control only by an *a priori* knowledge of its strength and by setting a proper guard-band between the coherent and IMDD combs. To this aim, in Fig. 3a we report the magnitude, obtained using the effects superposition of pump & probe results, of both XPM amplitude and phase noise varying the spectral occupancy of the IMDD pumps and their distance from a coherent probe as a means of system design. It should be also noted that this investigation looks at the XPM impairments before the DSP processing, whose action potentially masks the entity of the physical phenomenon due to the effectiveness of the CPE algorithm which is strongly implementation-dependent. Also, phase noise appears as a narrow-band effect (Fig. 3b) since it depends on the bandwidth of the IMDD pumps, thus slower with respect to the PM-QPSK symbol rate. It follows that common CPE algorithms should be able to track and compensate for it. However, it can trigger cycle-slips in the DSP depending on its intensity that may also vary with fiber polarization axes. The PDFs for the phase noise are plotted in Fig. 3c. In general, the guard-band must be set after characterizing the XPM phase noise, and verifying that the CPE is able to follow phase noise also in case of worst polarization case that randomly happens depending on the mechanical stress.

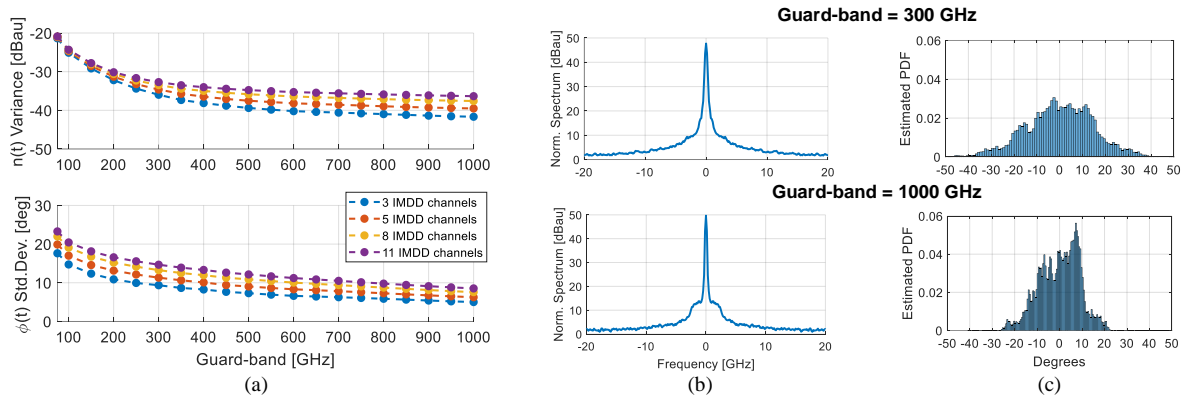

Fig. 3. (a) XPM intensity on PM-QPSK probe vs guard-band and number of IMDD pumps at $P_{pump} = 1$ dBm. (b) Phase noise $\phi(t)$ spectrum and (c) estimated pdf on PM-QPSK probe by 11 IMDD pumps at $P_{pump} = 1$ dBm at 300 GHz and 1 THz guard-band.

## 3. Comments and conclusions

We have observed via simulation the effects of 10G channels on 100G transmission and we have separated the amplitude from the phase noise effects. The amplitude noise coming from FWM and XPM can be managed by the SNR QoT figure together with the ASE noise and the 10G-to-100G NLI. On the other hand, the phase noise must be kept under control by setting a guard-band based on the maximum phase noise the DSP implementing the coherent receiver is able to tolerate. The knowledge of the different nature of 10G-to-100G effects can be also useful to develop specific DSP techniques aimed at mitigating the joint 10G/100G transmission.